\documentclass[letter]{aa}  
\usepackage[varg]{txfonts}
\usepackage[]{hyperref}

\begin{document} 

\title{Warped diffusive radio halo around the quiescent spiral edge-on galaxy NGC~4565}
\titlerunning{Diffusive radio halo around NGC~4565}

\author{
V.~Heesen\inst{1},
L.~Whitler\inst{2},
P.~Schmidt\inst{3},
A.~Miskolczi\inst{4},
S.~S.~Sridhar\inst{5},
C.~Horellou\inst{6},
R.~Beck\inst{3},
G.~G\"urkan\inst{7},
E.~Scannapieco\inst{2},
M.~Br\"uggen\inst{1},
G.~H.~Heald\inst{7},
M.~Krause\inst{3},
R.~Paladino\inst{8},
B.~Nikiel-Wroczy\'{n}ski\inst{9,10},
\and R.-J. Dettmar\inst{4}
}

\authorrunning{V. Heesen et al.}
\institute{University of Hamburg, Hamburger Sternwarte, Gojenbergsweg 112, D-21029 Hamburg, Germany\\
\email{volker.heesen@hs.uni-hamburg.de}
\and School of Earth and Space Exploration, Arizona State University, P.O. Box 871404, Tempe, AZ, 85287-1404, USA
\and Max-Planck-Institute f\"ur Radioastronomie, Auf dem H\"ugel 69, D-53121 Bonn, Germany
\and Astronomisches Institut der Ruhr-Universit\"at Bochum, Universit\"atsstr. 150, D-44780 Bochum, Germany
\and ASTRON, the Netherlands Institute for Radio Astronomy, Postbus 2, 7990AA, Dwingeloo, The Netherlands
\and Chalmers University of Technology, Dept of Space, Earth and Environment, Onsala Space Observatory, SE-439 92 Onsala, Sweden
\and CSIRO Astronomy and Space Science, PO Box 1130, Bentley WA 6102, Australia
\and INAF/Istituto di Radioastronomia, via Gobetti 101, I-40129 Bologna, Italy
\and Astronomical Observatory, Jagiellonian University, ul. Orla 171, 30-244 Krak\'ow, Poland
\and Leiden Observatory, Leiden University, Oort Gebouw, PO Box 9513, NL-2300 RA Leiden, the Netherlands}

\date{Received month DD, YYYY; accepted month DD, YYYY}

\abstract {Cosmic rays play a pivotal role in launching galactic winds, particularly in quiescently star-forming galaxies where the hot gas alone is not sufficient to drive a wind. Except for the Milky Way, not much is known about the transport of cosmic rays in galaxies.} 
{In this Letter, we present low-frequency observations of the nearby edge-on spiral galaxy NGC~4565 using the LOw-Frequency ARray (LOFAR). With our deep 144-MHz observations, we obtain a clean estimate of the emission originating from old cosmic-ray electrons (CRe), which is almost free from contamination by thermal emission.} 
{We measured vertical profiles of the non-thermal radio continuum emission that we fitted with Gaussian and exponential functions. The different profile shapes correspond to 1D cosmic-ray transport models of pure diffusion and advection, respectively.} 
{We detect a warp in the radio continuum that is reminiscent of the previously known H\,{\sc i} warp. Because the warp is not seen at GHz-frequencies in the radio continuum, its minimum age must be about 100~Myr. The warp also explains the slight flaring of the thick radio disc that can otherwise be well described by a Gaussian profile with an FWHM of 65\,arcsec ($3.7$\,kpc).}
{The diffusive radio halo together with the extra-planar X-ray emission may be remnants of enhanced star-forming activity in the past where the galaxy had a galactic wind, as GHz-observations indicate only a weak outflow in the last 40~Myr. NGC~4565 could be in transition from an outflow- to an inflow-dominated phase.}




\maketitle



\section{Introduction}

Radio haloes are a common feature around late-type spiral galaxies that are seen in an edge-on orientation \citep[e.g.][]{irwin_99a,wiegert_15a,heesen_18a}. The emission in the halo is dominated by the non-thermal radio continuum (synchrotron) component, which indicates cosmic ray electrons (CRe) and magnetic fields. This is corroborated by the detection of linearly polarised emission, which allows us to study the structure of magnetic fields in the halo that often display a characteristic X-shape in projection \citep[e.g.][]{tuellmann_00a,krause_09a,wiegert_15a}. The structure of the magnetic field is of importance for the transport of cosmic rays (protons, electrons, and heavier nuclei) from the disc into the halo. This is the scenario if cosmic rays are accelerated by diffusive shock acceleration in supernova (SN) remnants in the thin, star-forming galactic mid-plane \citep{bell_78a}. This connection is corroborated by the tight correlation between radio continuum luminosity and star formation rate (SFR) in galaxies \citep[e.g.][]{tabatabaei_17a}.


Cosmic ray transport can be described by two idealised modes,
advection and diffusion, and the superposition of both. This distinction is important because advection-dominated haloes are indicative of outflows and winds.  Outflows are directly observed over a wide range of galaxy masses, but in a smaller range of SFR surface densities.  Large outflows are ubiquitous in galaxies in which the SFR density per  unit area exceeds  $ \Sigma_{\rm SFR} \approx 0.1\,{\rm M}_\odot\,{\rm yr}^{-1}\,{\rm kpc}^{-2}$  \citep{heckman_03}, and they are more difficult to detect for lower rates \citep{chen_10}. 
Observations also show that discs with strong outflows are characterised by velocity dispersions between
$\sigma_{\rm \varv}^{\rm 1D} \approx $50--$100\,{\rm km\,s^{-1}}$ \citep[e.g.][]{genzel_11,swinbank_11}, which occur at high surface densities. At these velocity dispersions, a thermal runaway occurs that facilitates escape from the galaxy for supernova material \citep{scannapieco_12,scannapieco_13,sur_16}.


Advective haloes provide an avenue for detecting weaker outflows in
environments with low SFR surface densities. This interpretation is suggested by the correlation between advection and rotational velocity of a galaxy, where the latter is a proxy for the escape velocity \citep{heesen_18a,miskolczi_19a,schmidt_19a}. On the other hand, diffusive haloes allow us to measure diffusion coefficients and their energy dependencies, giving us measurements in addition to those in the Milky Way \citep{strong_07a}. Thus far, only one such example has been found in NGC~7462 \citep{heesen_16a}. This galaxy has an SFR surface density of $\Sigma_{\rm SFR}=1.6\times 10^{-3}\,\rm M_{\sun}\,yr^{-1}\,kpc^{-2}$, which is close to the threshold reported by \citet{rossa_03a} for the detection of extra-planar ionised gas (eDIG) in galaxies.





A threshold $\Sigma_{\rm SFR}$ for advective haloes (and thus,
galactic winds) is corroborated by a  possible threshold for
extra-planar X-ray emission, which is a direct tracer of the hot
SNe-heated gas. \citet{tuellmann_06a} found extra-planar X-ray
emission only in galaxies with $\Sigma_{\rm SFR}>3\times 10^{-3}\,\rm
M_{\sun}\,yr^{-1}\,kpc^{-2}$. If such a simple threshold exists, it
will also depend on the distribution of star formation in the disc; galaxies in which star formation is more concentrated, for instance,
nuclear starbursts, are more likely to have winds. Hydrodynamical simulations reported by \citet{vasiliev_19a} suggested a slightly higher SFR surface density for galaxy-wide outflows of $\Sigma_{\rm SFR}\approx 6\times 10^{-3}\,\rm
M_{\sun}\,yr^{-1}\,kpc^{-2}$, but the authors did not take the dynamical importance of cosmic rays into account.  Another key parameter
that will influence any outflow or wind is its mass surface density
\citep[MSD;][]{krause_18a}, where high MSDs are expected to suppress outflows.

In this Letter, we present first results from our survey of nearby
galaxies with the International LOw-Frequency ARray
\citep[LOFAR;][]{vanHaarlem_13a}, where we observed with the same setup as the
LOFAR Two-metre Sky Survey
\citep[LoTSS;][]{shimwell_17a}. Low-frequency radio continuum
observations have the benefit that we see the oldest CRe far away from
star formation sites in the halo and that the contamination from
thermal emission is minimal because the non-thermal spectrum rises steeply at low frequencies, whereas the thermal spectrum is essentially flat. 

NGC~4565 is an Sbc late-type spiral
galaxy at an assumed distance of $11.9$\,Mpc
\citep{tully_88a,radburn_smith_11a}, at an almost edge-on orientation
with an inclination angle of $87\fdg 5$ \citep{zschaechner_12a}, with
a low $\Sigma_{\rm SFR}$ of $0.74\times 10^{-3}\rm
M_{\sun}\,yr^{-1}\,kpc^{-2}$ \citep{wiegert_15a}. It has a total mass of $10^{11.4}\,\rm
M_{\sun}$, an MSD of $11.2\times 10^{7}\,\rm
M_{\sun}\,kpc^{-2}$, and an SFR of $0.73\,\rm M_{\sun}\,yr^{-1}$
\citep{wiegert_15a}. This means that NGC~4565 is a quiescent galaxy, with both
a low SFR and $\Sigma_{\rm SFR}$. It has an average MSD, in the middle of the range found in the sample of radio haloes studied by \citet{krause_18a}, who
  covered a range between $4$ and $16\times 10^{8}\,\rm
  M_{\sun}\,kpc^{-2}$. This work is a
low-frequency study of NGC~4565 complimentary to \citet[][]{schmidt_19a}, who presented the high-frequency ($1.5$ and 6\,GHz) view. Radio spectral indices $\alpha$ are defined as $S_{\nu}\propto\nu^{\alpha}$, where $S_{\nu}$ is the flux density and $\nu$ is the observing frequency.





\begin{figure}
        \includegraphics[width=\columnwidth]{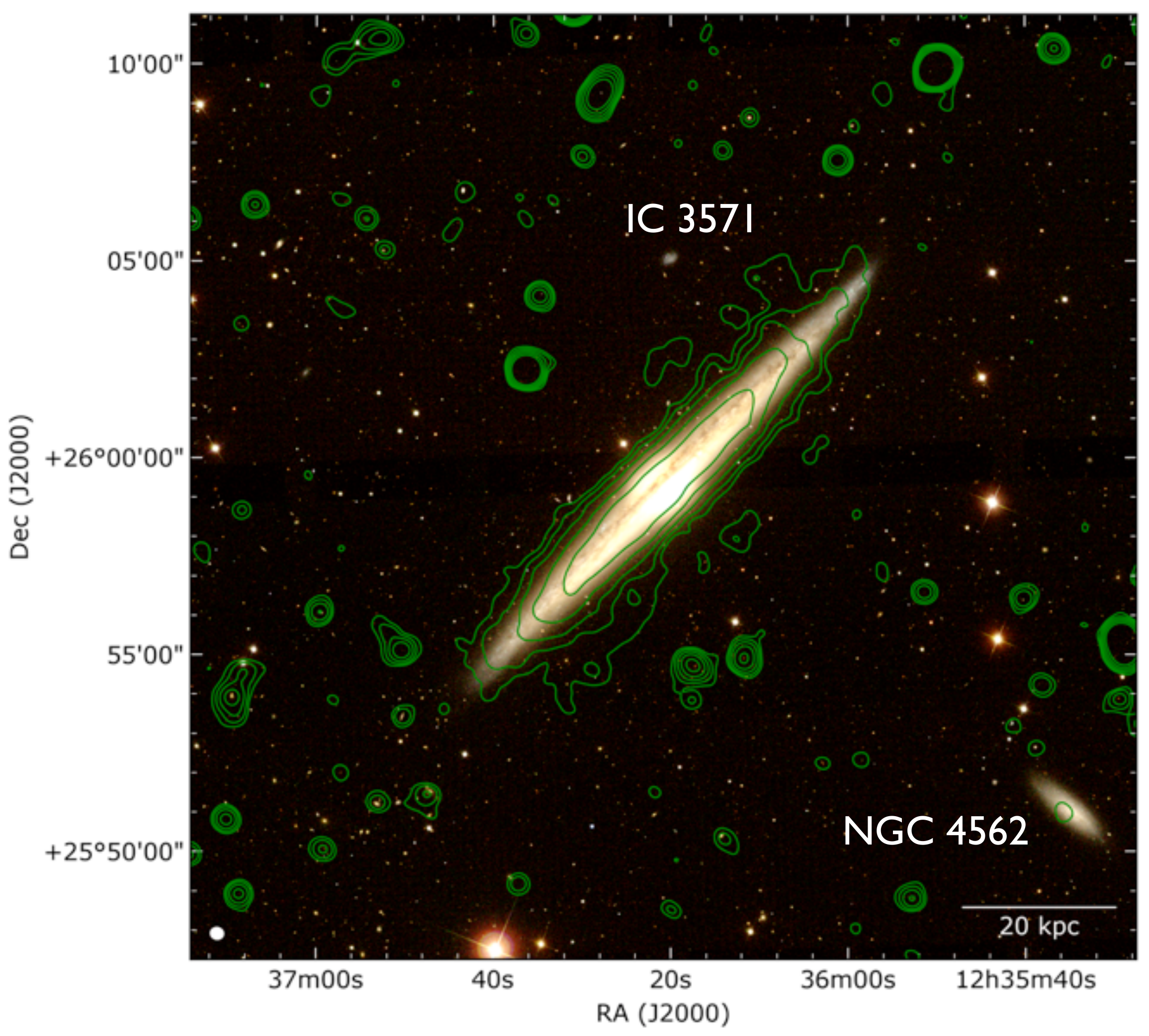}
    \caption{Radio continuum emission at 144\,MHz, presented as contours on an SDSS image, at $19.6\times 17.5\,\rm arcsec^2$ FWHM resolution. Contours start at a level of $3\sigma$ ($390\,\mu\rm Jy\,beam^{-1}$), increasing by a factor of 2. The size of the synthesised beam is shown in the bottom left corner in white.}
    \label{fig:144mhz_sdss}
\end{figure}

\section{Observations and data reduction}

Our High Band Antenna (HBA) LOFAR observations were taken on August
28 and October 12, 2017, as part of the LoTSS, where we observed the LoTSS pointing P194+27, which is $1\fdg 5$ away from our target. We used the HBA-dual
inner mode to conduct 8 hr observations, with the 48 MHz bandwidth
(120--168\,MHz), book-ended by 10 min flux-calibrator scans
(i.e.\ a total observing time  of $8.3$\,h). We stored the data at 16 channels per sub-band ($12.2\,\rm kHz$ frequency resolution) and at 1\,s time resolution. The data were reduced with the facet calibration technique, which mitigates the direction-dependent effects of the ionosphere and beam response that affect low-frequency radio continuum observations with aperture arrays, so that images close to the thermal noise level can be obtained \citep{van_weeren_16a,williams_16a}. Below we describe the data reduction in some more detail.



First, the $(u,v)$ data were calibrated with direction-independent methods using the {\small PREFACTOR} pipeline \citep{degasperin_19a}. Following the amplitude calibration using the flux densities of \citet{scaife_12a} for our primary calibrator, 3C~295, the station gain amplitudes and the phase variations due to the drift of the clocks and the Earth's ionosphere were determined. When they were determined, the instrumental calibration solutions were applied to the target data, which were then 
averaged to 10s time resolution and a frequency resolution of two-channels per sub-band (channel 
width of $97.656\,\rm kHz$). Then the data were calibrated in phase only using the global sky model~\citep[GSM;][]{scheers_11a}.  With the direction-independent calibration applied, the $(u,v)$ data were inverted and deconvolved with a wide-field {\small CLEAN} algorithm. As the final step of {\small PREFACTOR}, the {\small CLEAN} components of all the sources within the $8\degr$ field of view (FoV) were subtracted from the $(u,v)$ data.\footnote{\href{https://github.com/lofar-astron/prefactor}{https://github.com/lofar-astron/prefactor}}


\begin{figure*}
        \includegraphics[width=\textwidth]{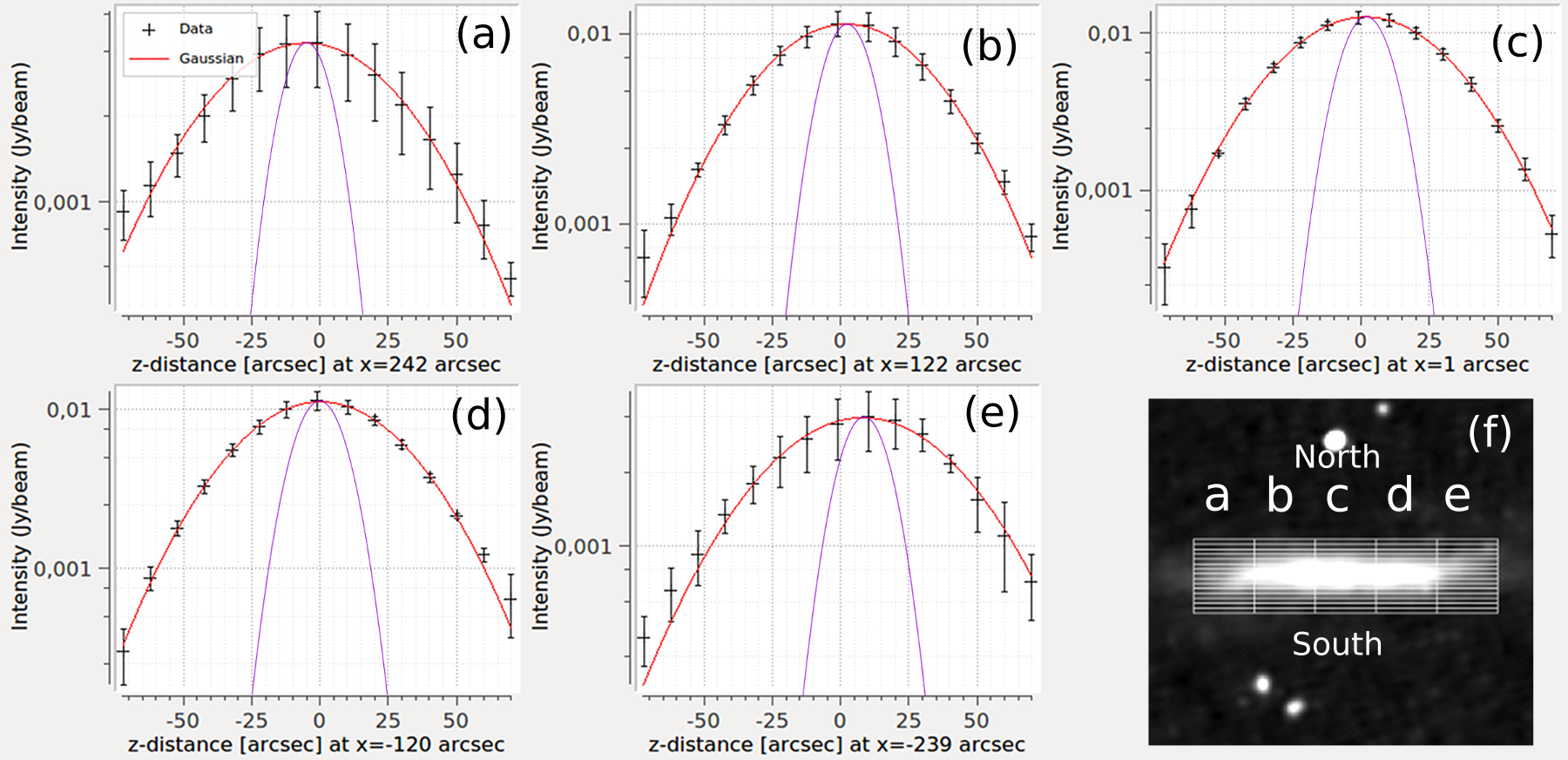}
    \caption{Vertical profiles of the non-thermal radio continuum
      emission at 144~MHz in
      strips at various offsets along the major axis [panels
      (a)--(e)] and the position of the strips [panel
      (f)]. Red solid lines show Gaussian model fits to the data, and
      purple solid lines show the Gaussian synthesised beam ($\rm
      FWHM=20\,arcsec$). Vertical tick marks are at $2.5\times$,
      $5\times$, and $7.5\times$ the next labelled tick value
      below. In the profiles, south is to the left and north is to the
    right.}
    \label{fig:profile}
\end{figure*}

For the direction-dependent calibration we used the {\small FACTOR} pipeline.\footnote{\href{https://github.com/lofar-astron/factor}{https://github.com/lofar-astron/factor}} The FoV was divided into approximately 20 facets around calibrator regions with integrated 167 MHz flux densities (of the full facet) in excess of $0.3$\,Jy.  Of these, facets in excess of $0.8$\,Jy were processed one at a time, beginning with the brightest facet. In the first step of the calibration, fast 10s phase solutions were determined in small chunks of $\approx$2\,MHz bandwidth to correct for the positional change and distortion of sources. In the second step, slow long-amplitude (tens of minutes) solutions were used to track the variation in the apparent flux density of a source. The target facets were corrected using the solution of a nearby facet.

The direction-dependent calibrated $(u,v)$ data were imported into the Common Astronomy Software Applications \citep[{\small CASA};][]{mcmullin_07a} and inverted and deconvolved with the MS--MFS {\small CLEAN} algorithm \citep{rau_11a}. We fitted for the frequency dependence of the skymodel ($\rm nterms=2$) and used angular scales of up to 88\,arcsec for the deconvolution. This is significantly smaller than the size of the galaxy (16\,arcmin), but we found that increasing the largest angular scale does result in large residual areas of negative flux densities above and below the galaxy. We verified that in spite of the small angular scale no significant flux was left in the residual map. We used Briggs weighting, setting the robust parameter to $0.2$, which resulted in a map with an effective central frequency of 144\,MHz with angular resolutions of 20\,arcsec FWHM with a map noise of $\sigma=130\,\mu\rm Jy\,beam^{-1}$. The radio continuum spectrum of this galaxy is discussed in Appendix~\ref{app:spectrum}.

\section{Results and discussion}
Figure~\ref{fig:144mhz_sdss} shows the radio continuum emission at
144\,MHz displayed as contours on an optical image from the Sloan Digital Sky Survey \citep[SDSS;][]{york_00a}. It is
striking that the emission is almost purely confined to the optical
disc with very little extra-planar emission. The extent of the first
contour along the major axis is 14\,arcmin, but the extent along
the minor axis is only $2.2$\,arcmin in the centre of the galaxy,
increasing to $2.9$\,arcmin south-east of the centre. This
galaxy therefore has a major-to-minor-axis ratio of around 5. This value can
be compared with the value reported by \citet{singal_15a}, who found in  their
sample of ten spiral edge-on galaxies a ratio of $1.4\pm 0.5$, similarly to \citet{wiegert_15a}, who stacked 30 of $1.5$-GHz images. In dwarf irregular galaxies, the ratio can even be lower than unity \citep{heesen_18b}. Clearly, the axis ratio of the contours depends on frequency, angular resolution, and sensitivity, but the stark contrast already shows that NGC~4565 is quite different from the galaxies that were studied so far.










\begin{table}
        \centering
        \caption{Best-fitting Gaussian and exponential one-component fits.}
        \label{tab:intensity_fits}
        \begin{tabular}{lcccc} 
                \hline
                Strip  & FWHM & Gauss & $h$ & Exp.\\
        & (arcsec) & $\chi^2_{\rm red}$ & (arcsec) & $\chi^2_{\rm red}$ \\\hline
                (a) & $71.87\pm 1.42$ & $0.37$ & $33.45\pm 1.96$ & $1.93$\\
                (b) & $64.04\pm 0.75$ & $0.48$ & $28.58\pm 1.78$ & $4.59$\\
                (c) & $61.41\pm 0.27$ & $0.34$ & $26.99\pm 2.17$ & $21.53$\\
                (d) & $60.94\pm 0.55$ & $0.43$ & $26.71\pm 1.79$ & $11.09$\\
                (e) & $70.42\pm 1.73$ & $0.34$ & $32.74\pm 2.33$ & $1.06$\\\hline
        \end{tabular}
\end{table}




A more quantitative way to measure radio halo properties is to use
vertical profiles of the non-thermal intensities, for which we have
subtracted the thermal emission from a combination of H $\alpha$ and
\emph{Spitzer} $24$ $\mu$m emission \citep[adopted from][]{schmidt_19a}.
This correction is small: the thermal fraction is only 2\%\ at
144\,MHz, with most of the emission concentrated in the thin
star-forming disc, so that the thermal fraction in the halo is
entirely negligible. In Fig.~\ref{fig:profile} we present these
profiles, which we measured with {\small BoxModels}
\citep{mueller_17a}, and fitted them with one-component Gaussian and exponential fits. The results are presented in Table~\ref{tab:intensity_fits}. It is immediately clear that they cannot be
fitted with one-component exponential profiles, except possibly in
the outer two profiles (strips a and e, because of their large error
bars), where we find $1.1<\chi_{\rm red}^2<21.5$. In contrast, the profiles can be
very well fitted by single-component Gaussian functions
($0.34<\chi^2_{\rm red}<0.43$), with an FWHM that increases from
$61.4$\,arcsec in the central stripe to $71.9$\,arcsec in the
north-western, outer stripe (strip e) and similarly in the
south-eastern, outer stripe (strip a). These measured extents are well
resolved (we recall the angular resolution of 20\,arcsec). We
therefore detect a slight flaring of the radio halo, as suggested by the shape of
the radio emission contours. This is also seen at higher frequencies \citep[][]{schmidt_19a}.
\begin{figure}
        \includegraphics[width=\columnwidth]{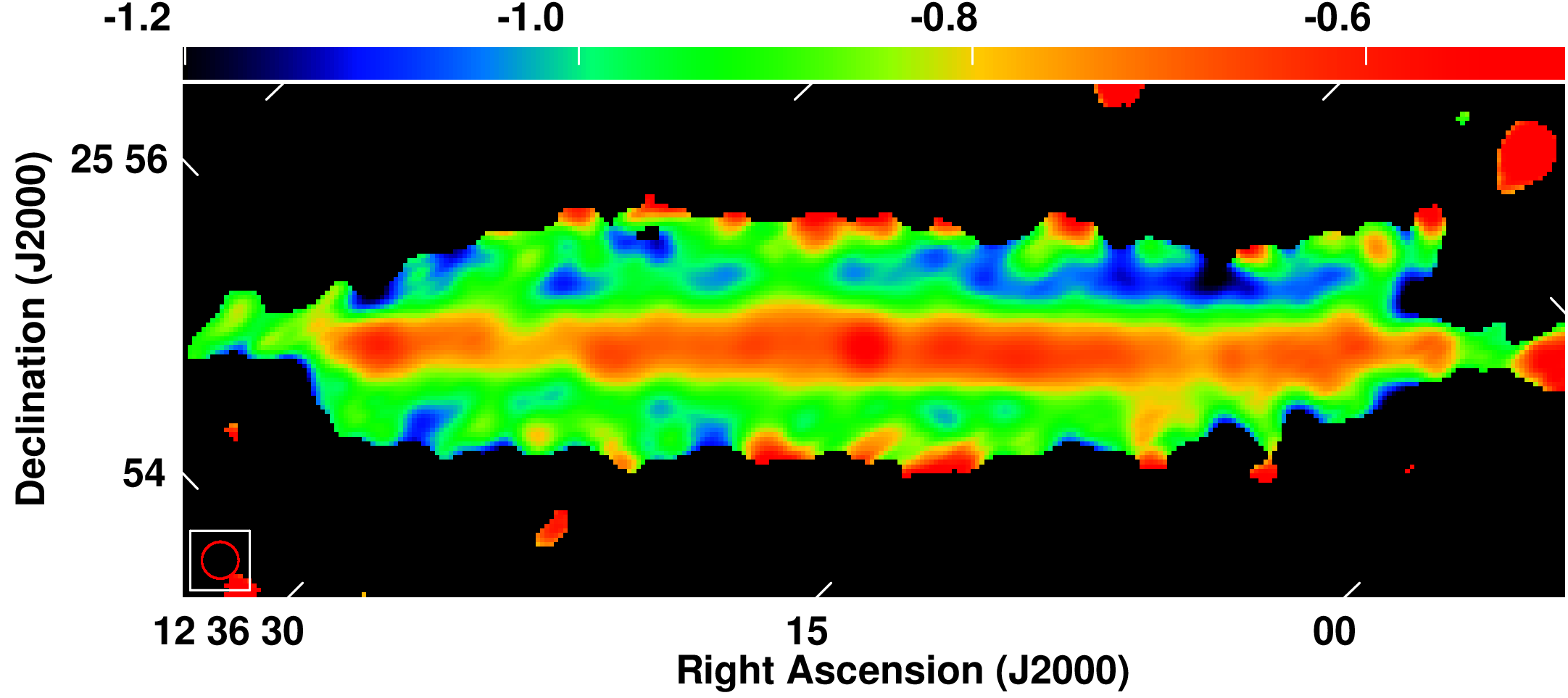}
    \caption{Non-thermal radio spectral index between 144 (LOFAR) and 1570\,MHz (JVLA) at 20\,arcsec FWHM resolution. The major axis has been rotated so that it is horizontal. The size of the synthesised beam is shown in the bottom left corner.}
    \label{fig:spix}
\end{figure}
\begin{table}
        \centering
        \caption{Best-fitting diffusion and advection models. Reduced $\chi^2$, diffusion coefficients ($D=D_0(E/{\rm GeV})^{\mu}$), and advection speeds ($V$) in the northern (N) and southern (S) haloes.}
        \label{tab:spinnaker}
        \begin{tabular}{lcccc} 
                \hline
                Strip  & LOFAR & JVLA & &\\
        & $\chi_{\rm red}^2$ & $\chi_{\rm red}^2$ & $D_0$ & $V$\\
        & (diff / adv) & (diff / adv) & ($10^{28}\,\rm cm^2\,s^{-1}$) & ($\rm km\,s^{-1}$)\\
                \hline
                (a) N & $0.05$ / $0.07$ & $0.38$ / $0.09$ & $1.5^{+8.5}_{-0.8}$ ($\mu=0.4$) & $66^{+180}_{-30}$ \\
                (a) S & $0.15$ / $0.24$ & $0.21$ / $0.08$ & $2.3^{+1.9}_{-1.0}$ ($\mu=0.0$) & $56^{+40}_{-20}$ \\
                (b) N & $0.74$ / $1.16$ & $1.50$ / $0.38$ & $3.1^{+1.7}_{-0.9}$ ($\mu=0.0$) & $82^{+30}_{-15}$ \\
                (b) S & $0.40$ / $0.92$ & $1.84$ / $0.54$ & $2.1^{+1.1}_{-0.6}$ ($\mu=0.0$) & $67^{+20}_{-10}$ \\
                (d) N & $1.55$ / $5.16$ & $6.96$ / $2.26$ & $1.5^{+0.9}_{-0.5}$ ($\mu=0.0$) & $54^{+30}_{-15}$ \\
        (d) S & $0.76$ / $3.31$ & $0.69$ / $0.13$ & $0.7^{+0.2}_{-0.1}$ ($\mu=0.5$) & $81^{+60}_{-20}$ \\
                (e) N & $0.64$ / $0.70$ & $0.63$ / $0.17$ & $3.3^{9.0}_{-1.6}$ ($\mu=0.0$) & $66^{+90}_{-25}$ \\
                (e) S & $0.04$ / $0.07$ & $0.15$ / $0.05$ & $1.6^{+2.5}_{-0.8}$ ($\mu=0.0$) & $46^{+50}_{-25}$ \\
                \hline
        \end{tabular}
\end{table}

For the first time, we can report a warp in the radio continuum (compare with
Fig.~\ref{fig:144mhz_sdss}) that is reminiscent of the H\,{\sc
  i} warp in its outer parts. The H\,{\sc i} warp can be seen in projection against the inner disc
\citep{zschaechner_12a}. The flaring of the radio halo
seems to be caused by the warp because the vertical intensity profiles
(with the exception of the central stripe) are asymmetric, in agreement
with the warp. We can therefore now give a minimum age for the warp of approximately 130~Myr, which is the spectral age of the CRe during which they are transported into the warp. The warp is not visible at GHz frequencies \citep{heesen_18a, schmidt_19a}, which means that we can rule out a younger age of 40~Myr or less. \citet{radburn_smith_14a} found young ($<600$~Myr) stars in the warp with a lack of older ($>1$~Gyr) stars, which places an upper limit on the age of the structure.

\begin{figure}
        \centering
        \includegraphics[width=0.9\columnwidth]{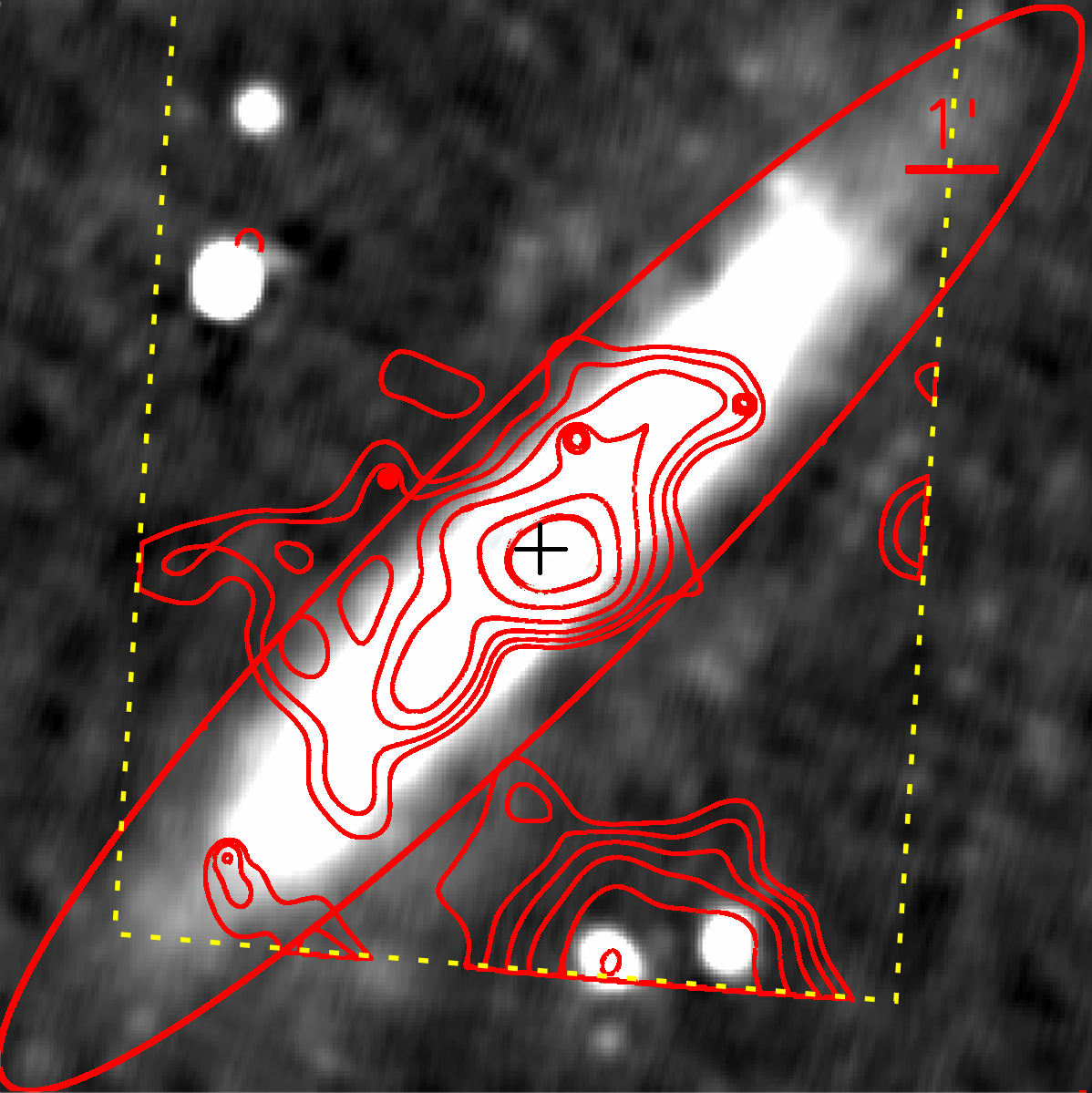}
    \caption{Soft X-ray emission ($0.5$--$1.5$\,keV) as contours \citep[][see their Fig.~4 for details]{li_13a}, overlaid on 144 MHz radio continuum emission as grey scale ($-0.6$ to $3\,\rm mJy\,beam^{-1}$) at 20\,arcsec FWHM resolution. The ellipse shows the $D_{25}$ optical diameter ($16.7$\,arcmin), the position of the nucleus is marked by a cross, and the yellow dashed line shows the \emph{Chandra} FoV.}
    \label{fig:144mhz_xray}
\end{figure}
We fitted the intensity profiles with 1D cosmic ray transport models for
advection and diffusion using {\small SPINNAKER} \citep{heesen_16a}. In order to calculate the radio spectral indices that are required as input, we
used a 1570 MHz map from $L$-band observations with the Karl G.\ Jansky Very
Large Array (JVLA), combined from the B, C, and D array \citep{schmidt_19a}. These data were taken as part of the `Continuum
Halos in Nearby Galaxies, an EVLA Survey'
\citep[CHANG-ES;][]{irwin_12a}. The resulting spectral index map is
shown in Fig.~\ref{fig:spix}. In the galactic mid-plane, relatively flat spectral indices $\alpha\approx -0.7$ are found, which is indicative of young CRe. In contrast,  the steep spectral indices of $\alpha\approx -0.9$ in the halo are indicative of old CRe. This is the expected behaviour if young CRe are injected at star formation sites in the mid-plane, from where they are transported into the halo. We therefore fitted the profiles except in the central stripe c, which might be influenced by an outflow from the Seyfert nucleus \citep{ho_97a}.


Our model assumes that the CRe are injected with a power law in the galactic mid-plane and lose their energy through synchrotron and inverse Compton losses. A steady state between injection and losses is then reached. We present the best-fitting parameters for these models in Table~\ref{tab:spinnaker} and the corresponding profiles in Appendix~\ref{app:models}. We found that pure CRe diffusion can describe our 144 MHz data better than advection, where the reduced $\chi^2$ is on average a factor of two smaller for diffusion. On the other hand, advection can describe the 1570-MHz profiles better than diffusion; the improvement is even more pronounced. This outcome can be explained if the galaxy is in the aftermath of a period with more intense star formation, as has been proposed by \citet{schmidt_19a}. The LOFAR map is then dominated by the CRe that were injected during a past starburst and had time to slowly diffuse away from star formation sites in the galactic mid-plane. In contrast, the emitting CRe at GHz frequencies are dominated by those that are advected in the current weak outflow.

This scenario is corroborated by several observations. The mid- and far-infrared emission from dust together with radiative transfer modelling suggest that the SFR has indeed declined in the past 100~Myr \citep{de_looze_12a}, which is in good agreement with the CRe lifetime at the LOFAR frequency. Furthermore, there is a reservoir of extra-planar X-ray emission in the halo that extends up to 10~kpc away from the galactic mid-plane (see Fig.~\ref{fig:144mhz_xray}). This emission is usually thought to be a remnant of a past starburst. At 1570 MHz, the vertical profiles of the non-thermal radio continuum can be described by two-component exponential functions with a thin-disc scale height of mostly only 20~pc \citep{schmidt_19a}; this means that the thin disc is much less extended than in other galaxies, where it has scale height of around 400~pc \citep{heesen_18a}. The data are consistent with either diffusion or advection in the halo (height $>$ 1~kpc), but with an outflow speed lower than the escape velocity. If an outflow currently exists, it is therefore a fountain flow rather than a wind.



We found mostly good fits to the data with the exception of strip d north, where neither diffusion nor advection leads to good fits. This strip is in the region of the H\,{\sc i} bridge that connects NGC~4565 with its satellite galaxy IC~3571 \citep[][see also Fig.~\ref{fig:144mhz_sdss}]{radburn_smith_14a}. It is therefore possible that tidal interaction governs the gas dynamics in this area and the transport of CRe with it, such that our idealised cosmic ray transport models no longer fit. We find diffusion coefficients that are in good agreement with \citet{schmidt_19a} and the canonical Milky Way value of $3\times 10^{28}\,\rm cm^2\,s^{-1}$ \citep{strong_07a}. We mostly failed to find any energy dependence, just as \citet{schmidt_19a} found at higher frequencies. While they were unable to formally distinguish between diffusion and advection, however, our new LOFAR data require diffusion to be the dominant transport process.



\section{Conclusions}

We have observed the nearby edge-on spiral galaxy NGC~4565 with LOFAR at 144\,MHz in the radio continuum to measure the distribution of CRe and magnetic fields. Our results are listed below.
\begin{itemize}
\item[] (i) The radio continuum emission is constrained to a very elongated thin distribution with a major-to-minor-axis ratio of around 5.
\item[] (ii) The vertical intensity profiles can be well fitted by a
  one-component Gaussian function, which is the first such a case at
  low frequencies. Our explanation is that NGC~4565 has an old ($\approx$100\,Myr) population of CRe that has slowly diffused away from the star-forming disc.
\item[] (iii) There are indications that NGC~4565 is in the aftermath of a period with more intense star formation. The weak outflow that may be seen at higher frequencies was therefore probably stronger in the past, which explains the diffusive low-frequency radio halo.
\end{itemize}


Outflows and inflows in galaxies are thought to occur in quasi-periodic fashion. Bursts of star formation lead to outflows that remove gas from the disc into the halo. This gas, in particular the warm component, will then fall back onto the disc as part of a so-called fountain flow during periods of less intense star formation \citep{chang_goo_18a}. NGC~4565 may be an example for a galaxy that we observe in transition state from an outflow to an inflow dominated period. This would explain why it behaves differently to NGC~7462, the only other galaxy in which we have discovered a diffusive radio halo so far. There, the GHz-profiles can be fitted by a one-component Gaussian fit, such that the galaxy would already be in the inflow-dominated phase and would entirely lack an outflow. 

When we visualize galactic winds in our minds, the local starburst galaxy M82 is arguably the most commonly used picture \citep[e.g.][]{leroy_15a}. However, the evidence is now mounting that outflows and winds play a crucial role in the evolution of even in fairly quiescent galaxies; for instance, the stellar population synthesis analysis of NGC~628 required the efficient removal of mass as a result of stellar feedback \citep{zaragoza-cardiel_19a}. Radio continuum observations with LOFAR open up a new window on the population of cosmic rays in these systems, allowing us to study their significance for galaxy evolution in much more detail.





{\small {\it Acknowledgements.} We would like to thank the anonymous referee for an insightful report. We also thank Hans-Rainer Kl\"ockner for carefully reading the manuscript. LOFAR, the Low Frequency Array designed and constructed by ASTRON, has facilities in several countries that are owned by various parties (each with their own funding sources) and that are collectively operated by the International LOFAR Telescope (ILT) foundation under a joint scientific policy. Funding for the Sloan Digital Sky Survey IV has been provided by the Alfred P. Sloan Foundation, the U.S. Department of Energy Office of Science, and the Participating Institutions. LW and ES were supported by the NSF under grant OSIE-1458445.}




\bibliographystyle{aa}
\bibliography{diff} 



\newpage

\appendix

\section{Radio continuum spectrum}
\label{app:spectrum}

\begin{figure}
        \centering
        \includegraphics[width=\columnwidth]{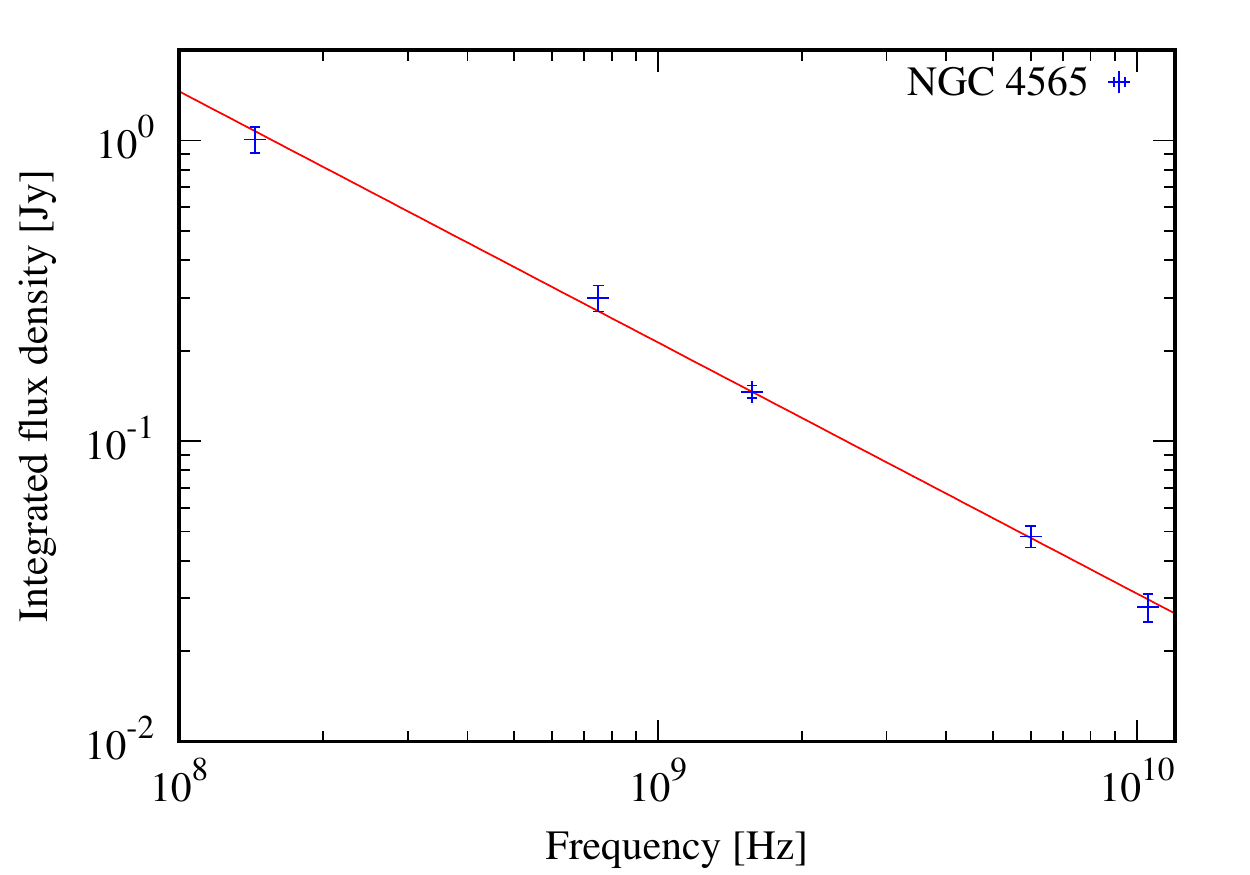}
    \caption{Radio continuum spectrum of NGC~4565. The integrated flux densities are shown, and the solid line is the best-fitting power law.}
    \label{fig:n4565_spectrum}
\end{figure}

In this appendix, we present the integrated radio continuum spectrum
of NGC~4565. We integrated the 144 MHz flux density in an area
approximately encompassing the 3$\sigma$ contour line, resulting in
$1.01\pm 0.10$\,Jy. The error of 10\%\ was adopted from
\citet{shimwell_17a}. This error is larger than the primary calibrator
model uncertainty of approximately 3\%\ \citep{scaife_12a} because the uncertainty in the beam model affects observations with
aperture arrays of fixed dipoles. Literature flux densities are presented in Table~\ref{tab:spectrum}. The radio continuum spectrum of NGC~4565 is therefore a power-law between 144\,MHz and $10.55$\,GHz with a spectral index of $-0.836\pm 0.022$ ($\chi_{\rm red}^2\equiv \chi^2/{\rm dof}=0.6$). The resulting radio continuum spectrum together with the best-fitting power law is presented in Fig.~\ref{fig:n4565_spectrum}.

\begin{table}
        \centering
        \caption{Integrated radio continuum flux densities.}
        \label{tab:spectrum}
        \begin{tabular}{lcc} 
                \hline
                $\nu$  & $S_{\nu}$ & Reference \\
        (GHz) & (Jy)&\\
        \hline
                $0.144$ & $1.01\pm 0.10$ & this work\\
                $0.750$ & $0.3\pm 0.03$ & \citet{heeschen_69a}\\
                $1.57$  & $0.146\pm 0.007$ & \citet{schmidt_19a}\\
                $6.0$   & $0.0482\pm 0.0039$ & \citet{schmidt_19a}\\
                $10.55$ & $0.028\pm 0.003$ & \citet{niklas_95a}\\
        \hline
        \end{tabular}
\end{table}

\section{Cosmic ray transport models}
\label{app:models}

In this appendix, we present the best-fitting advection and diffusion models for each strip. We assume that there is a steady state between CRe injection in the galactic mid-plane and CRe losses through synchrotron and inverse Compton radiation. As input parameters we used the equipartition magnetic field strength in the disc plane from \citet{schmidt_19a}. We used $U_{\rm IRF}/U_{\rm B}=0.35$ as the ratio of the energy densities of the interstellar radiation field (IRF) to that of the magnetic field. This means that synchrotron losses of the CRe dominate inverse Compton losses. We used {\small SPINNAKER} \citep{heesen_16a,heesen_18a} to calculate the 1D cosmic ray transport models for pure diffusion and advection with the fitting carried out in {\small SPINTERACTIVE} (A.\ Miskolczi 2018, private communication).\footnote{Both are available at https://github.com/vheesen/Spinnaker} 

The CRe were injected in the galactic mid-plane with a power-law $N=N_0 E^{-\gamma}$, where the CRe injection spectral index $\gamma$ is a free parameter. The magnetic field strength in the halo was parametrised by an exponential distribution in the halo, with $B(z)=B_0\exp(-z/h_{\rm B})$, where $z$ is the distance to the disc and $h_{\rm B}$ is the magnetic field scale height. The magnetic field scale height was fitted with the resulting scale heights presented in Table~\ref{tab:bfield}; values are between $2.5$ and $6.1$~kpc, where the best-fitting scale height is similar for diffusion and advection. We find that the scale heights in the different strips agree within the uncertainties. The maximum scale height, although still not significant when the errors are considered, is found in strip e (north), with the second highest scale height found in strip a (south). This is in agreement with the position of the warp that is visible in the radio continuum emission.

\begin{table}
        \centering
        \caption{Further parameters used during the modelling with {\scriptsize SPINNAKER}. The magnetic field strength in the galactic mid-plane, $B_0$, is an input. The magnetic field scale height $h_{\rm B}$ is fitted during the modelling, and so is the CRe injection spectral index $\gamma$.}
        \label{tab:bfield}
        \begin{tabular}{lccc} 
                \hline
                Strip  & $B_0$ & $h_{\rm B}$ & $\gamma$ \\
        & & (diff / adv) & (diff / adv)\\
        & ($\mu\rm G$) & (kpc) & \\
        \hline
                (a) N & $5.2$ & $3.4^{+2.2}_{-1.4}$ / $3.4^{+1.5}_{-1.4}$ & $2.6$ / $2.4$\\
                (a) S & $5.2$ & $4.9^{+3.0}_{-2.0}$ / $4.5^{+2.8}_{-1.7}$ & $2.5$ / $2.4$\\
                (b) N & $6.4$ & $3.5^{+0.7}_{-0.7}$ / $3.4^{+0.5}_{-0.5}$ & $2.5$ / $2.5$\\
                (b) S & $6.4$ & $2.9^{+0.5}_{-0.5}$ / $2.7^{+0.3}_{-0.3}$ & $2.4$ / $2.4$\\
                (d) N & $6.4$ & $3.6^{+1.0}_{-0.8}$ / $3.1^{+0.5}_{-0.5}$ & $2.5$ / $2.5$\\
        (d) S & $6.4$ & $3.2^{+0.5}_{-0.6}$ / $2.5^{+0.3}_{-0.3}$ & $2.4$ / $2.4$\\
                (e) N & $5.4$ & $6.0^{+6.0}_{-2.3}$ / $6.1^{+7.5}_{-3.0}$ & $2.7$ / $2.6$\\
                (e) S & $5.4$ & $3.6^{+2.0}_{-1.5}$ / $3.2^{+1.5}_{-1.2}$ & $2.6$ / $2.4$\\
                \hline
        \end{tabular}
\end{table}

\begin{figure*}
\centering  
\includegraphics[width=\textwidth]{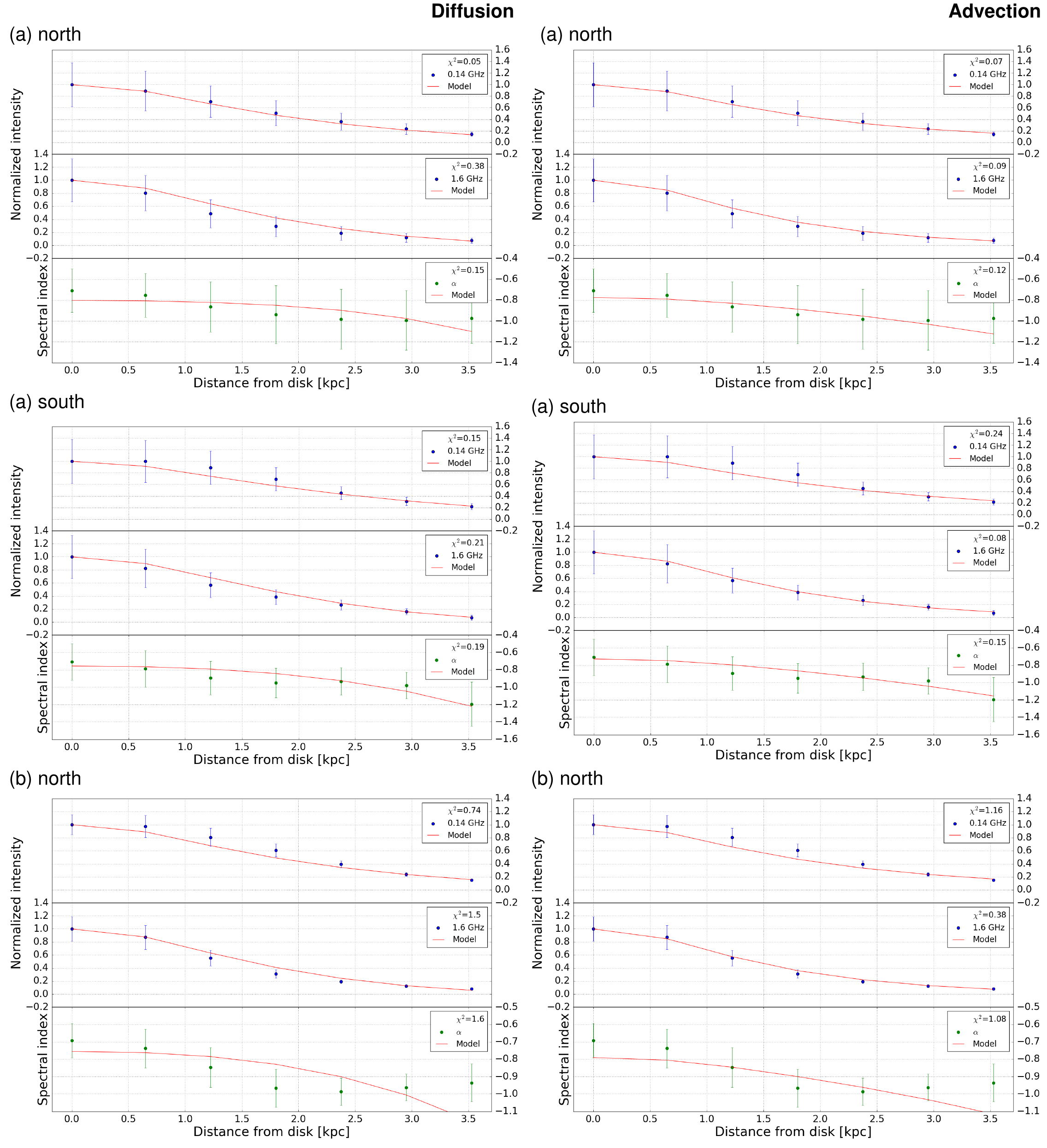}
    \caption{Best-fitting diffusion (\emph{left panels}) and advection (\emph{right panels}) models in strips a (north and south) and b (north). In each panel we show the vertical profiles. From top to bottom, we show the 144 MHz non-thermal intensities, the 1570 MHz non-thermal intensities, and the non-thermal radio spectral index. Solid lines show the best-fitting models calculated with {\scriptsize SPINNAKER}. In each panel we also present the reduced $\chi^2_{\rm red}$ for each observed profile and model profile.}
    \label{fig:fits1}
\end{figure*}

\begin{figure*}
\centering  
\includegraphics[width=\textwidth]{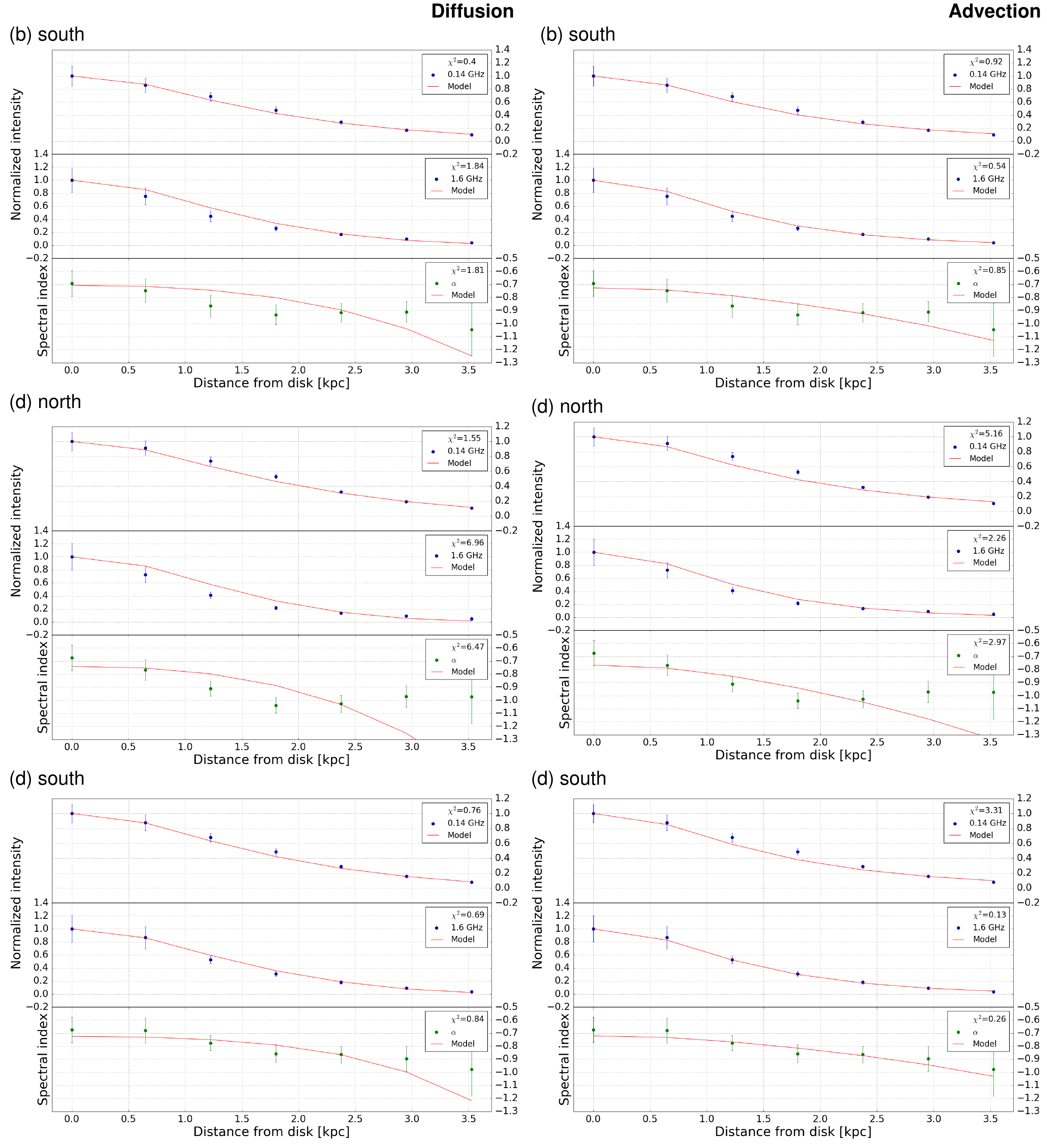}
    \caption{Best-fitting diffusion (\emph{left panels}) and advection (\emph{right panels}) models in strips b (south) and d (north and south). In each panel we show the vertical profiles. From top to bottom, we show the 144 MHz non-thermal intensities, the 1570 MHz non-thermal intensities, and the non-thermal radio spectral index. Solid lines show the best-fitting models calculated with {\scriptsize SPINNAKER}. In each panel we also present the reduced $\chi^2_{\rm red}$ for each observed profile and model profile.}
    \label{fig:fits2}
\end{figure*}

\begin{figure*}
\centering  
\includegraphics[width=\textwidth]{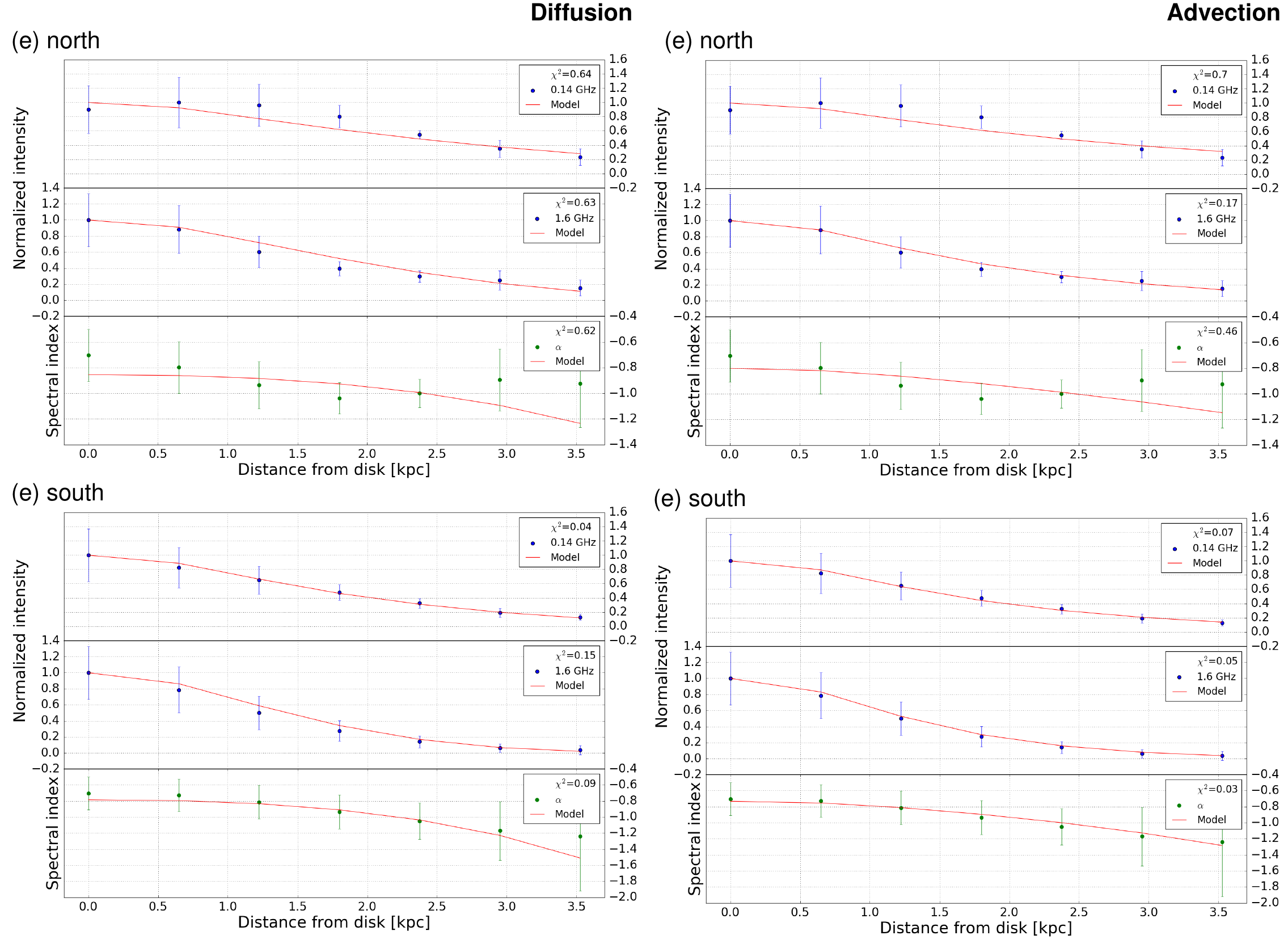}
    \caption{Best-fitting diffusion (\emph{left panels}) and advection (\emph{right panels}) models in strip e (north and south). In each panel we show the vertical profiles. From top to bottom, we show the 144 MHz non-thermal intensities, the 1570 MHz non-thermal intensities, and the non-thermal radio spectral index. Solid lines show the best-fitting models calculated with {\scriptsize SPINNAKER}. In each panel we also present the reduced $\chi^2_{\rm red}$ for each observed profile and model profile.}
    \label{fig:fits3}
\end{figure*}

In Figs~\ref{fig:fits1}--\ref{fig:fits3} we show the best-fitting models in the four strips in which we fitted the models (excluding the central strip). We treated the northern and southern haloes separately, therefore we fitted eight different profiles each with pure diffusion and advection, so that there are 16 different panels. 

\end{document}